\documentclass[apj]{emulateapj}
\usepackage{natbib}
\usepackage{graphicx}
\usepackage{epsfig}

\newcommand{\fesc}{\ifmmode{f_{\rm esc}}\else{$f_{\rm esc}$}\fi}
\newcommand{\fescs}{\ifmmode{f_{\rm esc}^\star}\else{$f_{\rm esc}^\star$}\fi}
\newcommand{\kms}{\ifmmode{{\rm km~s^{-1} }}\else{km s$^{-1}$}\fi}
\newcommand{\cubecm}{\ifmmode{{\rm cm^{-3} }}\else{cm$^{-3}$}\fi}
\newcommand{\lsim}{\lower0.3em\hbox{$\,\buildrel <\over\sim\,$}}
\newcommand{\gsim}{\lower0.3em\hbox{$\,\buildrel >\over\sim\,$}}

\newcommand{\Ms}{\ifmmode{M_\odot}\else{$M_\odot$}\fi}

\newcommand{\tvir}{\ifmmode{T_{\rm{vir} }}\else{$T_{\rm{vir}}$}\fi}
\newcommand{\mvir}{\ifmmode{M_{\rm{vir} }}\else{$M_{\rm{vir}}$}\fi}
\newcommand{\rvir}{\ifmmode{r_{\rm{vir} }}\else{$r_{\rm{vir}}$}\fi}

\newcommand{\jj}{\ifmmode{J_{21}}\else{$J_{21}$}\fi}
\newcommand{\flw}{\ifmmode{F_{LW}}\else{$F_{LW}$}\fi}

\begin{document}

\shorttitle{FLOWS AROUND MERGING BLACK HOLE BINARIES}
\shortauthors{VAN METER ET AL.}

\title{Modeling Flows Around Merging Black Hole Binaries} 

\author{James R. van Meter\altaffilmark{1}, 
  John H. Wise\altaffilmark{2},
  M. Coleman Miller\altaffilmark{3}, 
  Christopher S. Reynolds\altaffilmark{3},
  Joan Centrella\altaffilmark{1}, 
  John G. Baker\altaffilmark{1}, 
  William D. Boggs\altaffilmark{1},\altaffilmark{4},
  Bernard J. Kelly\altaffilmark{1},
  and Sean T. McWilliams\altaffilmark{1}}

\altaffiltext{1}{Gravitational Astrophysics Laboratory, NASA Goddard
  Space Flight Center, Greenbelt, MD 21114}
\altaffiltext{2}{Laboratory for Observational Cosmology, NASA Goddard
  Space Flight Center, Greenbelt, MD 21114}
\altaffiltext{3}{University of Maryland, Department of Astronomy,
  College Park, MD 20742}
\altaffiltext{4}{University of Maryland, Department of Physics
  College Park, MD 20742}
\email{james.r.vanmeter@nasa.gov}

\begin{abstract}

  Coalescing massive black hole binaries are produced by the mergers of
  galaxies.  The final stages of the black hole coalescence produce
  strong gravitational radiation that can be detected by the
  space-borne LISA.  In cases where the black hole merger takes
  place in the presence of gas and magnetic fields, various types of
  electromagnetic signals may also be produced.  Modeling such
  electromagnetic counterparts of the final merger requires evolving
  the behavior of both gas and fields in the strong-field regions
  around the black holes.  We have taken a step towards solving this problem
  by mapping the flow of pressureless matter in the dynamic, 3-D
  general relativistic spacetime around the merging black holes. We
  find qualitative differences in collision and outflow speeds,
  including a signature of the merger
  when the net angular momentum of the matter is low,
  between the results from single and
  binary black holes, and between nonrotating and rotating holes in
  binaries.  If future magnetohydrodynamic results confirm these
  differences, it may allow assessment of the properties of the
  binaries as well as yielding an identifiable electromagnetic
  counterpart to the attendant gravitational wave signal.
  
\end{abstract}


\section{Introduction}

Electromagnetic signatures of the coalescence of two supermassive
black holes would, in concert with the detection of gravitational
waves from the event, allow a high-precision determination of both the
redshift and the luminosity distance to the merger.  This would
provide a precise probe of cosmology, limited mainly by uncertainties
in the magnification, and hence the luminosity distance, due to weak
gravitational lensing \citep{2003CQGra..20S..65H}.  As a result, there
has been substantial recent interest in mechanisms that could produce
electromagnetic emission from a surrounding accretion disk
\citep{2002ApJ...567L...9A,2005ApJ...622L..93M,2006MNRAS.372..869D,
  2006ApJ...637...27K,2007APS..APR.T7002P,2007APS..APR.S1010B,2008ApJ...682..758S,
  2008ApJ...676L...5L,
  2008ApJ...684..835S,2008PhRvL.101d1101K,2008ApJ...684..870K,
  Haiman:2009te,O'Neill:2008dg,2009CQGra..26i4032H,
  Phinney_decadal,2009arXiv0906.0825C,Megevand:2009yx}.
However, there have been relatively few investigations of the
electromagnetic signatures that could emerge from the dynamic
spacetime near the binary during its last few orbits (for a recent
exception, see \citealt{2009arXiv0905.1121P}).

There are reasons to believe that if significant gas exists
around the binary in its last few orbits, then intense but 
short-lived electromagnetic signals will be produced.  For
example, since the binary orbital speed can be half the speed
of light or more, even in the Newtonian case the slingshot mechanism could
be particularly effective in ejecting matter or producing collisions
that have a high Lorentz factor.  In addition, it has been
proposed that magnetic fields in the vicinity of the orbit
could be wound up and amplified to such a degree that they would
produce a strong Poynting outflow (R. Blandford, personal
communication).  Exploration of such effects would require
simulations that involve magnetohydrodynamics (MHD) in a dynamical
spacetime.

Here we take a step towards such simulations by tracing the paths of
pressureless test particles (i.e., particles of non-zero rest mass in
geodesic motion) in the dynamical spacetime of coalescing black holes.
Our goal is to determine whether the outflows and collisions are
sufficiently high-speed, compared with the equivalents around a single
black hole, that unique signatures of the binary motion seem
plausible.  This will pave the way for the extensive development
required to do a full analysis of MHD in a dynamical spacetime.  In
\S~2 we discuss our simulation setup, which uses a code that has been
used to model successfully the coalescence of black holes of different
masses, spin magnitudes, and orientations.  We choose units in which
$c=1$ and $G=1$; with this, the total mass $M$ is related to time and
distance by $M \sim 5 \times 10^{-6}(M/M_{\odot})\; {\rm s} \sim
1.5(M/M_{\odot})\; {\rm km}$.  In \S~3 we present and discuss our
results, in which we examine first the coalescence of equal-mass
nonspinning black holes and then show that equal-mass black holes with
aligned spins produce significantly more extreme effects.

\section{Simulation Setup}

In the following simulations the spacetime was computed by
solving Einstein's field equations using finite differencing methods of 
numerical relativity \citep{Imbiriba:2004tp,Baker:2008mj}.  
On the initial slice, the black holes are
represented by ``punctures" \citep{Brandt:1997tf} and the constraint equations
are solved using a multigrid solver \citep{Brown:2004ma}.  The black holes
are then evolved forward in time by integrating the time-dependent
Einstein equations given by the BSSN formulation \citep{Shibata:1995we,Baumgarte:1998te}.
We use coordinate conditions that allow the black holes to move
freely across the grid \citep{Baker:2005vv,Campanelli:2005dd,vanMeter:2006vi}.  
Meanwhile the test matter is modeled
by non-interacting point-particles.  To evolve the test matter, we
integrate the geodesic equation for each particle in this spacetime.
The metric at the location of each particle is obtained by interpolation
from the computational grid.

The geodesic code was tested for a variety of trajectories around a
single black hole, both spinning and nonspinning.  For coordinate
conditions in which the conversion to Boyer-Lindquist coordinates was
known, the particle worldline was compared with independent
calculations and found to be at least 4th-order convergent in the
grid-spacing, as were the constants of motion, energy ($E$), angular
momentum ($L$), and Carter's constant ($Q$).  For the coordinate
conditions closer to those of our typical binary black hole runs, the
constant of motion $L$, readily calculable in any axisymmetric
coordinate system, was still found to be at least 4th-order
convergent.

Initial data for the spacetime background used here was chosen to
represent two black holes, each of mass $m = M/2$, in nearly circular inspirals.  In
our simulations the black holes were either spinless, or were
each given a spin of $a/m = 0.8$, aligned with the orbital angular
momentum.  The initial linear momenta of the black holes were given by
the post-Newtonian approximation, and the initial separation of the
black holes was chosen so as to give at least five orbits
before merger.
As a control case, we have also run equivalent simulations with a single isolated
nonrotating black hole of mass $M$.

  In these spacetimes, approximately 75,000 geodesic
particles were initially distributed uniformly throughout a solid
annulus of inner radius $8M$, outer radius $25M$, and vertical full
thickness $10M$.  We excluded particles within the inner radius to
avoid transient signatures from particles initially near the
horizons.  We have chosen such a geometrically thick disk because
these are the disks that potentially have high enough inward radial
speeds to keep up with the shrinking binary, as opposed to being
stranded at large radii as the binary coalesces (see
\citealt{2005ApJ...622L..93M}).  

We explored two initial velocity configurations for the particles.
In the first configuration (``orbital''), the initial velocities
are randomly distributed around a tangential
velocity $V_c$ that would give a circular orbit in a Schwarzschild
spacetime of mass $M$, resulting in a scale height of $5M$.
In the
second configuration (``isotropic''), we consider an extreme case where the particles only
have random velocities, with each component sampled from a Gaussian
distribution of standard deviation $V_c/\sqrt{3}$.

We output the positions and proper 4-velocities of the particles every
$\sim 1M$ and $\sim 0.5M$ in the nonrotating and rotating cases,
respectively.  Particles within a horizon are discarded.  We use the
relative velocities between two ``colliding'' particles as an energy
estimate of the material within the disk.  We define a particle
collision as two particles traveling within $r_{\rm c} = 0.1M$ of
each other, assuming that they travel on straight lines between successive 
output times.  We have found that the collision energies are not sensitive
to $r_{\rm c}$ in the range of $0.01M$ to $1M$.  For every particle, we
search the nearest 8 neighbors for any particle that has a closest
approach less than $r_{\rm c}$, which we mark as a collision.

With this list of collisions, we can now obtain the Lorentz factor of
each colliding particle, in locally Minkowskian coordinates, as
follows.  Given two colliding particles $A$ and $B$, first we compute
the scalar product of the proper 4-velocity $u_A^{\mu}$ of particle
$A$ with the proper 4-velocity $u_B^{\mu}$ of particle $B$.  Note that
regardless of the original coordinates in which the product is
computed, it exactly equals the negative of the Lorentz factor of
particle $A$ in the rest-frame of particle $B$ (as it would be
calculated in locally Minkowskian coordinates).  That is,
\begin{equation}
u_A^{\mu}u_B^{\nu}g_{\mu\nu}=u_A^{\mu'}u_B^{\nu'}g'_{\mu\nu}=-u_A^{0'}
\end{equation}
because $u_B^{\mu'}=\delta_0^{\mu}$ and $g_{\mu\nu}'=\eta_{\mu\nu}$,
where the primes refer to the inertial frame of particle $B$.  We then
transform to the center-of-momentum frame by solving for the Lorentz
factor $\gamma_{com}$ characterizing the Lorentz transformation matrix
that transforms $u_A^{\mu'}$ and $u_B^{\mu'}$ to $u_{A(com)}^{\mu}$
and $u_{B(com)}^{\mu}$ such that
\begin{eqnarray}
u_{A(com)}^0=u_{B(com)}^0=\gamma_{com}, \; u_{A(com)}^i=-u_{B(com)}^i.
\end{eqnarray}
We finally obtain:
\begin{equation}
\gamma_{com}=\sqrt{\frac{1}{2}(1-u_A^{\mu}u_B^{\nu}g_{\mu\nu})}.
\end{equation}

\section{Results and Discussion}

\begin{figure} [t]
  \plotone{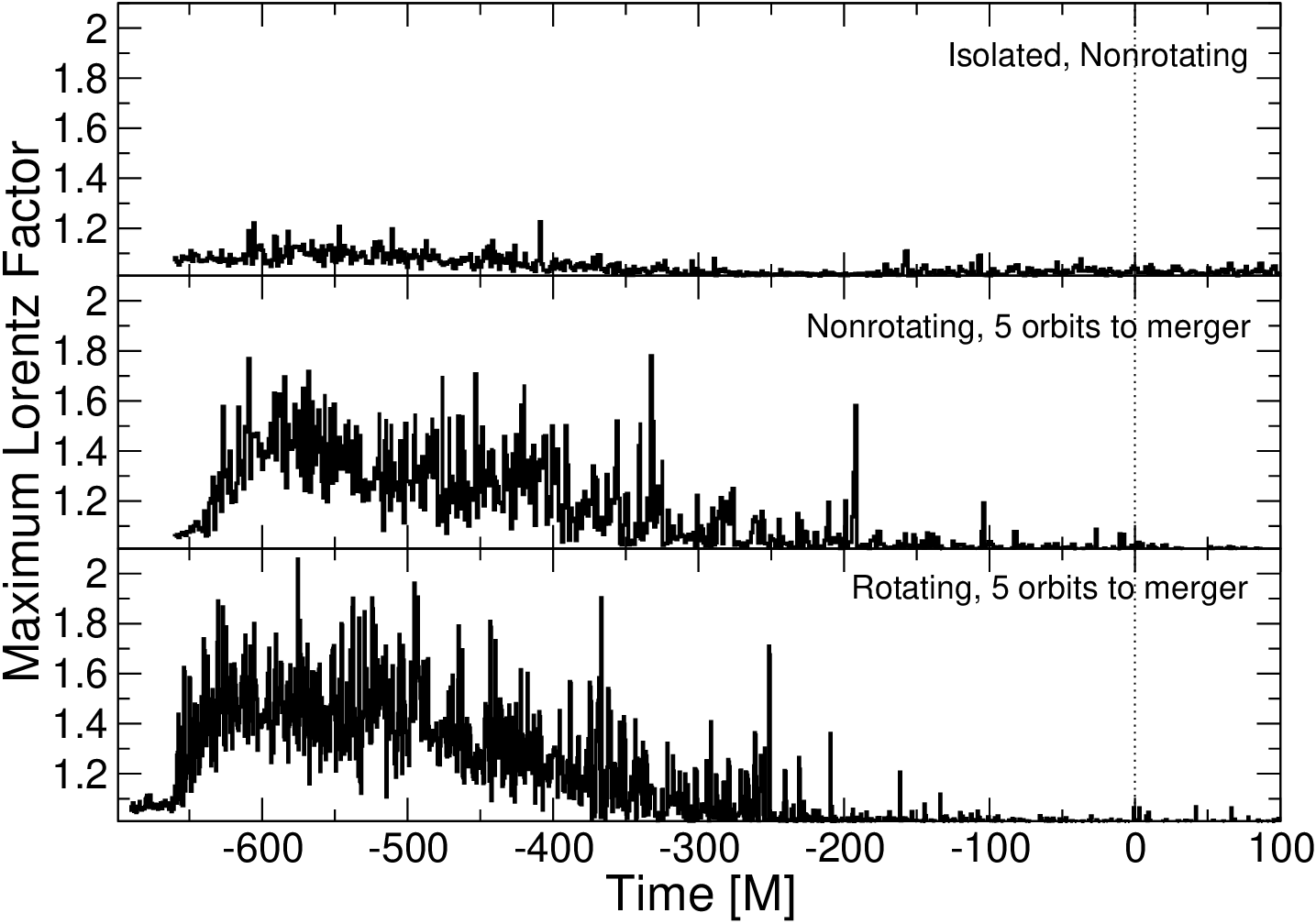}
  \caption{Maximum Lorentz factors of collisions between particles as
    function of time before merger for initial velocities that include
    an orbital velocity plus a random isotropic component.  The three
    panels are for the following cases: (top) isolated nonrotating
    black hole, (middle) two spinless black holes, merging in approximately five
    orbits, (bottom) two black holes each with spin parameters of
    $a/m=0.8$, merging in approximately six orbits.  In all cases both black holes
    have mass $m$.  The data have been aligned with respect to the time axis such
    that the common apparent horizon is first detected at $t=0$.  $M$ is the total
    system mass. 
    }
  \label{fig:max_gamma_disk}
\end{figure}

\begin{figure} [t]
  \plotone{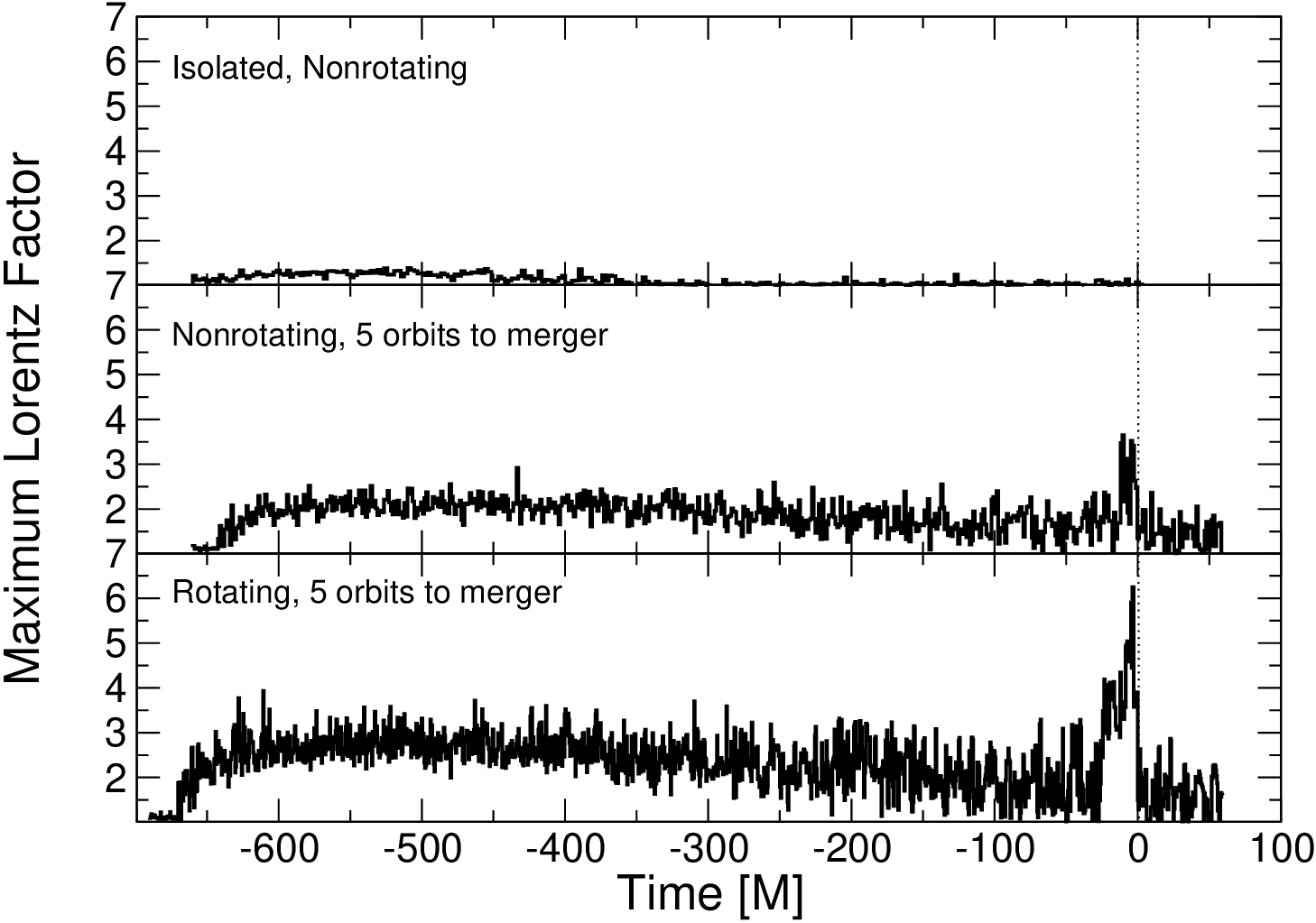}
  \caption{Same as Figure \ref{fig:max_gamma_disk} but for particles
    whose initial velocity is purely random and isotropic, with no net
    orbital component. From this figure we note that (a)~in both
    binary cases there is a clear maximum in Lorentz factor just prior
    to merger, whereas (as expected) the isolated case has a steady
    Lorentz factor, and (b)~in the spinning case, the collisions
    throughout and near merger are more violent than in the
    nonrotating runs.}
  \label{fig:max_gamma_therm}
\end{figure}

\begin{figure} [t]
  \epsscale{0.8}
  \plotone{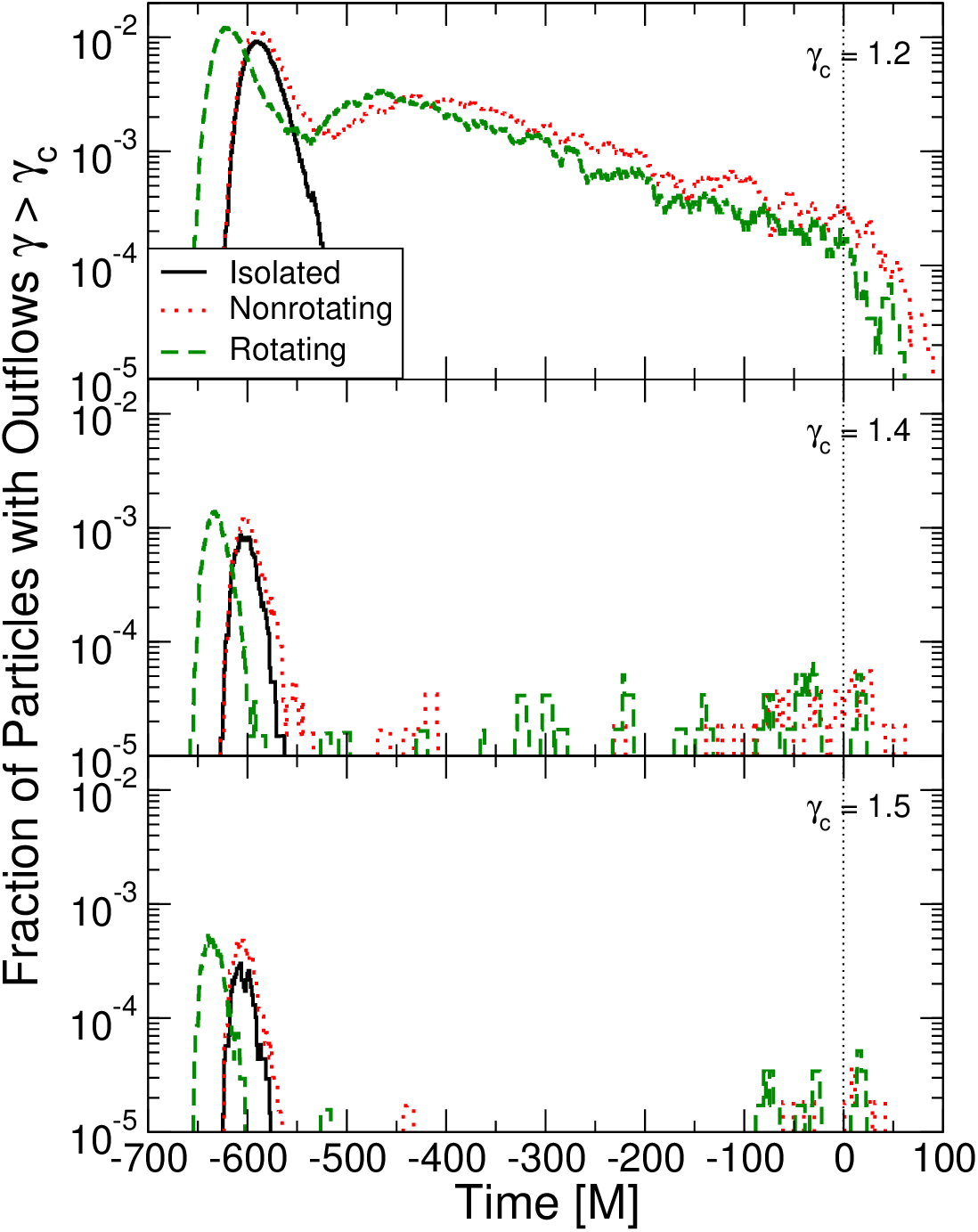}
  \caption{The fraction of particles with positive radial velocities
    in a shell with surfaces at $25M$ and $30M$ with a corresponding
    Lorentz factor greater than (top to bottom) 1.2, 1.4, 1.5,
    comparing the cases with an isolated spinless black hole (solid
    black), a binary with spinless equal-mass black holes (dotted
    red), and a binary with equal-mass black holes with spins of $a/m
    = 0.8$ (dashed green).  On the scale plotted, one particle gives a
    fractional value of 1/75,000 or $1.3 \times 10^{-5}$.}
  \label{fig:outflow_disk}
\end{figure}

\begin{figure} [t]
  \epsscale{0.8}
  \plotone{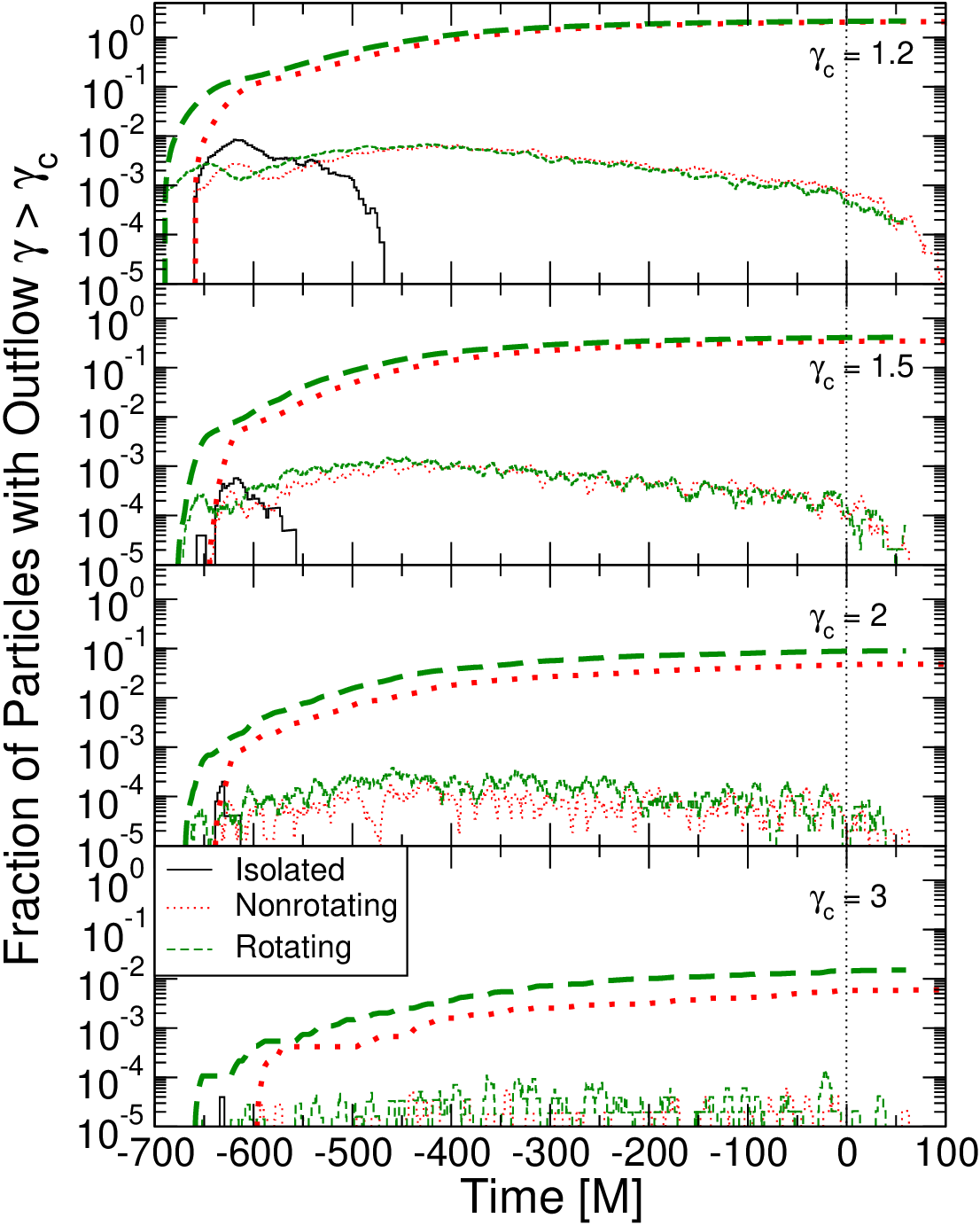}
  \caption{Same as Figure \ref{fig:outflow_disk} but for the case of
    particles having only initial random velocities and Lorentz
    factors greater than (top to bottom) 1.2, 1.5, 2, 3.  
    Also shown are the cumulative sums of these fractions.}
  \label{fig:outflow_therm}
\end{figure}

Our main results are displayed in Figures \ref{fig:max_gamma_disk}
through \ref{fig:outflow_therm}.  In Figures \ref{fig:max_gamma_disk}
and \ref{fig:max_gamma_therm}, with initially orbital and isotropic particle velocities
respectively, we show the maximum Lorentz factor of a collision
between any two particles as a function of time, for several different
setups including one with an isolated Schwarzschild black hole for
comparison.  Because collisions do not affect particle trajectories in
our calculations (i.e., the geodesic paths are unaltered), we expect
these to be upper bounds on the true Lorentz factors of collisions
given that in the MHD situation realized by nature there will be
dissipation of relative speeds through collisions.

In  Figures \ref{fig:max_gamma_disk} and \ref{fig:max_gamma_therm},
the top panel shows the control case, in which the particles orbit
around a solitary black hole. The other two panels involve nonspinning
(middle panel) and spinning (bottom panel) binaries. The merger in
these cases, defined by the detection of a common apparent horizon,
takes place at $t=0$.

Our control case shows as expected that after a short initialization phase the
maximum Lorentz factor is roughly constant.  Here the highest-speed
collisions are between particles that have both fallen close to the
horizon and thus have acquired speeds considerably in excess of their
initial orbital speeds.  Given that the particles have a ``thermal''
velocity distribution in which the random speed is comparable to the
orbital speed, a reasonable fraction of particles is captured by the
black hole.

	In the binary runs the varying accelerations caused by the binary motion
	are expected to alter significantly the energy and angular momentum of
	particles within $\sim 2$ binary orbital radii, just as in the
	Newtonian slingshot analogue.  More rapid orbits therefore are
	expected to lead to higher-speed collisions and (as is evident in
	Figure~\ref{fig:outflow_therm}) faster outflows, with the effect
	maximized shortly before merger because this is when the orbital
	speeds are greatest.

	Examining Figure~\ref{fig:max_gamma_therm} we find that these expectations
	are largely met for the isotropic particle velocities.
	There is indeed a strong peak in the maximum collision Lorentz factor
	just before merger.  This peak is highest when the holes are rotating,
	as is reasonable because for the aligned spin configuration the holes
	can get closer and thus orbit faster before they plunge together.  The
	Lorentz peaks are in all cases considerably higher than in the
	isolated hole runs, which might lead us to cautious optimism that this
	phase might be observably different for a binary merger than for an
	isolated hole.  However, in the case of particles with initial average
	velocities that are approximately orbital, the central region of the
	disk remains evacuated near the time of merger, as the particles have
	found quasi-stable orbits further out.  In this case there is no
	discernible merger signature, as seen in
	Figure~\ref{fig:max_gamma_disk}.  We conjecture that the inclusion of
	viscosity would push orbiting matter closer towards the binary,
	resulting in energetics between the above two extremes, and recovering
	a merger signature.  If this is confirmed by future detailed MHD
	calculations, it may provide another way to identify an
	electromagnetic counterpart to the gravitational waves from a binary
	supermassive black hole merger.

	Other features of these figures are, we believe, artifacts of the
	limitations of our simulation.  For example, even with isotropic
	particle velocities, in both the spinning and the nonspinning binary runs we
	see an overall decrease in the maximum Lorentz factor of collisions
	during inspiral.  We suspect that this is because our setup includes no
	angular momentum transport between particles, hence as the simulation
	progresses we run out of particles that would naturally plunge towards
	the holes and produce very high speed collisions.

	In Figure \ref{fig:outflow_disk} and \ref{fig:outflow_therm} we
	explore the outflow Lorentz factor, measured in a shell of thickness
	$5M$ with its inner surface equal to initial disk outer radius of
	$25M$.  Particularly in the case of isotropic particle velocities, because we
	have a significant random speed in our thick disks, the isolated hole
	has some fraction of particles that move rapidly outward, but we see
	that the fraction of such particles is (as expected) very small
	compared to those of the binary runs where particles are flung out via
	a slingshot effect.

	The main meaningful difference, in Figure~\ref{fig:outflow_therm}, is
	between the spinning and nonspinning binary
simulations.  At a given time before merger, due to spin-orbit
coupling in the spinning case, the holes are orbiting faster than in
the nonspinning case.  As a result, the slingshot effect is stronger
at all times from merger, and near merger itself the speeds are higher
yet.  It is also possible that frame-dragging due to spin plays a more
direct role, by increasing the ejection speeds in a manner akin to an
eggbeater.  At the highest speeds seen, there is no parallel for the
isolated black hole run.  Depending greatly on the nature of the
interactions of the outflow (e.g., the development of shocks or the
initiation of Fermi acceleration), it is possible that high energy
radiation such as gamma-rays may be produced in ways that make it
possible to discriminate between single and binary black holes, or
even possibly between slowly-rotating and rapidly-rotating black
holes.

In summary, our exploratory calculations represent a step towards
realistic simulations of the accretion flow near binary supermassive
black holes close to merger.  Under certain conditions our results
show differences between single and binary black holes, and between
nonrotating and rapidly rotating binary black holes, which is
encouraging for future observations.  As such, they motivate more
realistic hydrodynamic and magnetohydrodynamic calculations, which
will be our next steps.  If some of the qualitative differences found
in our test particle calculations (with isotropically distributed velocities)
are maintained in
future work with fluids, this will imply prompt electromagnetic
signatures that can be correlated with gravitational radiation signals
detected by the space-based Laser Interferometer Space Antenna
\citep{LISASciCase,Lang:2008gh}, allowing precise cosmological probes and
constraints on dark energy \citep{Arun:2008xf} and even testing
fundamental principles such as the relative speed of propagation of
photons and gravitational waves.

\acknowledgments

We performed these calculations on Discover at NASA / GSFC and
Pleiades at NASA / AMES.  CSR and MCM acknowledge partial support from
the National Science Foundation under grant AST 06-07428.  MCM was
also supported in part by NASA ATP grant NNX08AH29G.  The work at
Goddard supported in part by NASA grant 06-BEFS06-19.  JHW, STM and BJK were
supported by appointments to the NASA Postdoctoral Program at the
Goddard Space Flight Center, administered by Oak Ridge Associated
Universities through a contract with NASA.

\clearpage

\bibliography{refs}

\end{document}